\documentclass{article}
\usepackage[letterpaper]{geometry}

\usepackage{epsfig}
\usepackage{graphicx}
\usepackage{color}
\usepackage[acmtitlespace,nocopyright]{OneClmn}
\usepackage{times}
\usepackage{amsmath}
\usepackage{amssymb}
\usepackage{xspace}
\usepackage{enumerate}

\usepackage[cmtip,arrow]{xy}
\usepackage{pb-diagram,pb-xy}

\newcommand {\mm}[1] {\ifmmode{#1}\else{\mbox{\(#1\)}}\fi}

\newcommand{\denselist}{\itemsep 0pt\parsep=1pt\partopsep 0pt}

\newcommand{\proof}{\noindent{\sc Proof.~}}
\newcommand{\eop}{\hfill\usebox{\smallProofsym}\bigskip}  %
\newsavebox{\smallProofsym}                            
\savebox{\smallProofsym}                               %
{
\begin{picture}(6,6)                                   %
\put(0,0){\framebox(6,6){}}                            %
\put(0,2){\framebox(4,4){}}                            %
\end{picture}                                          %
}                                                      %
\makeatletter
\long\def\@makecaption#1#2{%
  \vskip\abovecaptionskip
  \sbox\@tempboxa{\small #1: #2}%
  \ifdim \wd\@tempboxa >\hsize
    \small #1: #2\par
  \else
    \global \@minipagefalse
    \hb@xt@\hsize{\hfil\box\@tempboxa\hfil}%
  \fi
  \vskip\belowcaptionskip}
\makeatother

\newcommand{\Aspace}        {\mm{{\mathbb A}}}
\newcommand{\Bspace}        {\mm{{\mathbb B}}}
\newcommand{\Cspace}        {\mm{{\mathbb C}}}
\newcommand{\Dspace}        {\mm{{\mathbb D}}}
\newcommand{\Espace}        {\mm{{\mathbb E}}}

\newcommand{\Rspace}        {\mm{{\mathbb R}}}

\newcommand{\Vspace}        {\mm{{\mathbb V}}}
\newcommand{\Xspace}        {\mm{{\mathbb X}}}
\newcommand{\Tspace}        {\mm{{\mathbb T}}}
\newcommand{\Uspace}        {\mm{{\mathbb U}}}
\newcommand{\Yspace}        {\mm{{\mathbb Y}}}

\newcommand{\Bgroup}        {\mm{\sf B}}
\newcommand{\Cgroup}        {\mm{\sf C}}
\newcommand{\Dgroup}        {\mm{\sf D}}
\newcommand{\Egroup}        {\mm{\sf E}}

\newcommand{\Ggroup}        {\mm{\sf G}}
\newcommand{\Hgroup}        {\mm{\sf H}}

\newcommand{\Tgroup}        {\mm{\sf T}}
\newcommand{\Tgrouptoo}     {\mm{\sf T\!}}
\newcommand{\Ugroup}        {\mm{\sf U}}
\newcommand{\Vgroup}        {\mm{\sf V}}
\newcommand{\Wgroup}        {\mm{\sf W}}

\newcommand{\Region}[1]     {\mm{\mathcal W}_{{#1}}}
\newcommand{\lRegion}[1]    {\mm{\lambda}_{{#1}}}
\newcommand{\rRegion}[1]    {\mm{\varrho}_{{#1}}}
\newcommand{\Bcal}          {\mm{\mathcal B}}
\newcommand{\Pcal}          {\mm{\mathcal P}}
\newcommand{\Vcal}          {\mm{\mathcal V}}

\newcommand{\bmap}          {\mm{\sf b}}
\newcommand{\cmap}          {\mm{\sf c}}
\newcommand{\dmap}          {\mm{\sf d}}
\newcommand{\emap}          {\mm{\sf e}}
\newcommand{\fmap}          {\mm{\sf f}}
\newcommand{\gmap}          {\mm{\sf g}}
\newcommand{\hmap}          {\mm{\sf h}}
\newcommand{\jmap}          {\mm{\sf j}}
\newcommand{\tmap}          {\mm{\sf t}}

\newcommand{\umap}          {\mm{\sf u}}
\newcommand{\Ddgm}[2]       {\mm{\rm Dgm}_{#1}{({#2})}}
\newcommand{\Odgm}[2]       {\mm{\rm Ord}_{#1}{({#2})}}
\newcommand{\Rdgm}[2]       {\mm{\rm Rel}_{#1}{({#2})}}
\newcommand{\Edgm}[2]       {\mm{\rm Ext}_{#1}{({#2})}}

\newcommand{\rank}[1]       {\mm{\rm rank\,}{#1}}
\newcommand{\kernel}[1]     {\mm{\rm ker\,}{#1}}
\newcommand{\cokernel}[1]   {\mm{\rm cok\,}{#1}}
\newcommand{\image}[1]      {\mm{\rm im\,}{#1}}

\newcommand{\htop}          {\mm{h_{\rm top}}}
\newcommand{\hbot}          {\mm{h_{\rm bot}}}
\newcommand{\htopinv}       {\mm{h^{-1}_{\rm top}}}
\newcommand{\hbotinv}       {\mm{h^{-1}_{\rm bot}}}

\newcommand{\Maxdist}[2]    {\mm{\|{#1}-{#2}\|}_\infty}
\newcommand{\Pdist}[2]    {\mm{\|{#1}-{#2}\|}_{\Pcal}}
\newcommand{\capsp}         {{\; \cap \;}}
\newcommand{\cupsp}         {{\; \cup \;}}

\newtheorem{result}{}

\newcommand{\tot}           {\leftrightarrow}
\newcommand{\Tot}           {\longleftrightarrow}
\newcommand{\spans}[1]      {\langle #1 \rangle}

\title{Homology and Robustness \\~
       of Level and Interlevel Sets
       \thanks{This research is partially supported by the National
               Science Foundation (NSF) under grant DBI-0820624
               and the Defense Advanced Research Projects Agency (DARPA)
               under grants HR0011-05-1-0057 and HR0011-09-0065.}
       }

\author{Paul Bendich\thanks{IST Austria (Institute of Science and
            Technology Austria), Kloster\-neu\-burg, Austria, and Department of Mathematics, Duke University, Durham, North Carolina.}, 
        Herbert Edelsbrunner\thanks{IST Austria (Institute of Science
            and Technology Austria), Kloster\-neu\-burg, Austria,
            Departments of Computer Science
            and of Mathematics, Duke University, Durham, North Carolina,
            and Geomagic, Research Triangle Park, North Carolina.},
        Dmitriy Morozov\thanks{Departments of Computer Science and of
            Mathematics, Stanford University, Stanford, California.} and
        Amit Patel\thanks{Geometrica, INRIA-Saclay, Orsay, France,
            and IST Austria (Institute of Science and
            Technology Austria), Kloster\-neu\-burg, Austria.}
}

\begin{document}
\maketitle

\begin{abstract}
 Given a function $f: \Xspace \to \Rspace$ on a topological space,
 we consider the preimages of intervals and their homology groups
 and show how to read the ranks of these groups from the extended
 persistence diagram of $f$.
 In addition, we quantify the robustness of the homology classes
 under perturbations of $f$ using well groups, and we show
 how to read the ranks of these groups from the same extended
 persistence diagram.
 The special case $\Xspace = \Rspace^3$ has ramifications
 in the fields of medical imaging and scientific visualization.
\end{abstract}

\vspace{0.1in}

\section{Introduction}
\label{sec1}

The work reported in this paper has two motivations,
one theoretical and the other practical.
The former is the recent introduction of \emph{well groups} in the study of mappings
$f: \Xspace \to \Yspace$ between topological spaces.
Assuming a metric space of perturbations, we have such
a group for each subspace $\Aspace \subseteq \Yspace$,
each bound $r \geq 0$ on the magnitude of the perturbation,
and each homological dimension $p$.
These groups, and the diagrams that they generate,
extend the boolean concept of transversality
to a real-valued measure we refer to as \emph{robustness}.
Using this measure, we can quantify the robustness of a fixed point
of a mapping \cite{EMP10} and prove the stability of the apparent contour
of a mapping from an orientable $2$-manifold to $\Rspace^2$ \cite{EMP11}.
In this paper, we contribute to the general understanding
of well groups by studying the real-valued case. Along the way, we also extend the general theory of well
groups to incorporate relative well groups.
Specifically,
\begin{enumerate}\denselist
  \item[I.] we give a general definition of relative well groups given
    a mapping $f: \Xspace \to \Yspace$, a number $r \geq 0$, and a nested pair
    $\Aspace' \subseteq \Aspace$ of subspaces of $\Yspace$, and
  \item[II.]  we characterize the relative well groups of $f: \Xspace \to \Rspace$
    whenever $\Aspace$ is an interval and $\Aspace'$ is a subset of the endpoints.
\end{enumerate}
Applications of this theoretical work are anticipated in medical imaging
and scientific visualization,
where data in the form of real-valued functions is common.
To mention one example, it is common to acquire information about
internal organs through a magnetic resonance image,
which results in a $3$-dimensional array of intensity values,
best viewed as a function from the unit cube to the real line.
The predominant method for highlighting or extracting relevant substructures
of this image uses preimages of real values.
Generically, these are $2$-manifolds, commonly referred to as
\emph{contours} or \emph{isosurfaces} \cite{KOBPS97}.
Sometimes, these $2$-manifolds are complemented by preimages of intervals,
referred to as \emph{interval volumes} in visualization \cite{FMS95}.
In this paper, we call the preimage of a value a \emph{level set},
and the preimage of an interval an \emph{interlevel set},
in which the interval can be closed, open, or half-open.
We contribute to the state-of-the-art by
\begin{enumerate}\denselist
  \item[III.]  explaining how the homology of level and interlevel sets
    can be read off the extended persistence diagram of the function, and
  \item[IV.]   describing how the robustness of features in
    level and interlevel sets, quantified through well groups,
    can be read off the same diagram.
\end{enumerate}
Our results add up to a `point calculus' in algebraic topology
for mining the rich homological information contained in the extended
persistence diagram of a  real-valued function.
The compactness of the data representation and the
efficiency of the mining operations make the diagram
an attractive graphical interface tool for studying $3$-dimensional images.
We view this tool as complementary to the contour spectra described in
\cite{BPS97}, which plot continuously varying quantities,
such as area and volume, across the family of level sets.
The most novel aspect of our diagram is the robustness information,
which has previously not been available.
This novelty is combined with the unprecedented ease with which homological
information is accessible.
There is also evidence for the practicality of the interface
provided by the fast oct-tree implementation of
the described concepts \cite{BEK10},
which has been used to study $3$-dimensional images of root systems
of agricultural plants.

\paragraph{Outline.}
In Section \ref{sec2}, we review necessary background on persistence,
zigzag modules, and well groups.
In Section \ref{sec3}, we explain the point calculus for interlevel sets.
In Section \ref{sec4}, we extend the point calculus to include
the robustness information provided by the well groups.
Finally, Section \ref{sec5} concludes the paper with a brief discussion
of the contributions and of future research directions.

\section{Background}
\label{sec2}

We divide the background material into three parts,
introducing persistence and extended persistence in Section \ref{sec21},
explaining the extension to zigzag modules and level set pyramids
in Section \ref{sec22},
and defining absolute and relative well groups in Section \ref{sec23}.

\subsection{Forward Maps}
\label{sec21}

Traditional persistent homology is based on a nested sequence of spaces,
which induces a linear sequence of homology groups connected by maps
from left to right.
We describe this concept in two steps.

\begin{figure}[hbt]
 \vspace*{0.1in}
 \centering
 \resizebox{!}{1.4in}{\input{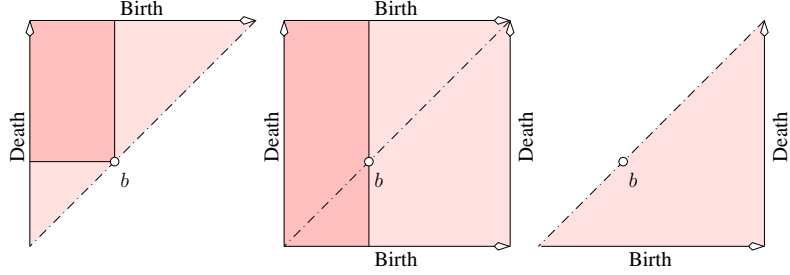}}
 \caption{From left to right:  the ordinary, extended, and relative
   subdiagrams of $\Ddgm{}{f}$.
   The number of points (not shown) in the dark shaded regions is equal
   to the rank of the homology group of the sublevel set defined by $b$.}
\label{fig:subdiagrams}
\end{figure}

\paragraph{Persistence.}
The persistence of homology classes along a filtration of a topological space
can be defined in a quite general context \cite{EdHa08}. 
For this paper, we need only a particular type of filtration, one defined
by the sublevel sets of a tame function.
Given a real-valued function $f$ on a compact topological space $\Xspace$,
we consider the filtration of $\Xspace$ via the \emph{sublevel sets}
$\Xspace_r(f) = f^{-1} (-\infty,r] $, for all real values $r$. 
Whenever $r < s$, the inclusion $\Xspace_r(f) \hookrightarrow \Xspace_s(f)$
induces maps on the homology groups
$\Hgroup_p(\Xspace_r(f)) \to \Hgroup_p(\Xspace_s(f))$, for each dimension $p$.
Here we will use field coefficients so that the homology groups are torsion-free and
are therefore vector spaces over the field.
Often we will suppress the homological dimension from our notation,
writing $\Hgroup(\Xspace_r(f)) = \bigoplus_p \Hgroup_p(\Xspace_r(f))$;
in this case, we will always assume that all maps $\Hgroup(\Xspace_r(f)) \to \Hgroup(\Xspace_s(f))$
decompose into the direct sum of maps on each factor.
A real value $r$ is called a \emph{homological regular value} of $f$
if there exists $\epsilon > 0$ such that the inclusion
$\Xspace_{r-\delta}(f) \hookrightarrow \Xspace_{r+\delta}(f)$ induces
an isomorphism between homology groups for every $\delta < \epsilon$. 
If $r$ is not a homological regular value, then it is a 
\emph{homological critical value}. 
We say that $f$ is \emph{tame} if it has finitely many
homological critical values and if the homology groups
of each sublevel set have finite rank.
Assuming that $f$ is tame, we enumerate
its homological critical values $r_1 < r_2 < \ldots < r_n$.
Choosing $n+1$ homological regular values $s_i$
such that $s_0 < r_1 < s_1 < \ldots <r_n < s_n$,
we put $\Xspace_i = \Xspace_{s_i}(f)$.
The inclusions $\Xspace_i \hookrightarrow \Xspace_j$ induce
maps $\fmap^{i,j}: \Hgroup(\Xspace_i) \to \Hgroup(\Xspace_j)$ for
$0 \leq i \leq j \leq n$ and give the following filtration:
\begin{equation}
  0 = \Hgroup(\Xspace_0)  \to  \Hgroup(\Xspace_1)  \to  \ldots
                          \to  \Hgroup(\Xspace_n) = \Hgroup(\Xspace) .
  \label{eqn:persfilt}
\end{equation}
We say a class $\alpha \in \Hgroup(\Xspace_i)$
is \emph{born} at $\Xspace_i$ if $\alpha \not \in \image{\fmap^{i-1,i}}$.
A class $\alpha$ born at $\Xspace_i$ is said to \emph{die entering}
$\Xspace_j$ if $\fmap^{i,j}(\alpha) \in \image{\fmap^{i-1,j}}$
but $\fmap^{i,j-1}(\alpha) \not \in  \image{\fmap^{i-1,j-1}}$.
We remark that if a class $\alpha$ is born at $\Xspace_i$,
then every class in the coset
$[\alpha] = \alpha + \image{\fmap^{i-1,i}}$ is born at the same time.
Of course, whenever such an $\alpha$ dies entering $\Xspace_j$,
the entire coset $[\alpha]$ also dies with it.
We represent $[\alpha]$ graphically as the point $(r_i, r_j)$ in the plane.
Drawing all birth-death pairs as points, we get diagrams like the ones
sketched in Figures \ref{fig:subdiagrams} and \ref{fig:twopants}.
Supposing that $b \in \Rspace$ is different from all homological critical values,
we collect all points in the upper-left quadrant defined by $(b,b)$
to get all classes born before $b$ and still alive;
see the left diagram in Figure \ref{fig:subdiagrams}.
Their number is the rank of the homology group of the sublevel set,
$\rank{\Hgroup (\Xspace_b (f))}$.

Observe that we really need the extended plane to draw the points
because some classes get born but never die, so the corresponding points have
$\infty$ as their second coordinates.
There is an elegant way around this minor annoyance, which we now describe.

\paragraph{Extended persistence.}
Since the filtration in (\ref{eqn:persfilt}) begins with the zero group
but ends with a potentially nonzero group, it is possible to have classes
that are born but never die.
We call these \emph{essential} classes, as they represent
the actual homology of the space $\Xspace$.
To measure the persistence of the essential classes, we
follow \cite{CEH09} and extend the sequence in (\ref{eqn:persfilt})
using relative homology groups. 
More precisely, we consider for each $i$
the \emph{superlevel set} $\Xspace^i = f^{-1} [s_{n-i},\infty)$.
Note that we have $\Xspace^0 = \emptyset$ and $\Xspace^n = \Xspace$ by compactness.
For $i < j$, the inclusion $\Xspace^i \hookrightarrow \Xspace^j$
induces a map on relative homology
$\Hgroup(\Xspace,\Xspace^i) \to \Hgroup(\Xspace,\Xspace^j)$.
These maps therefore give rise to the following extended filtration:
\begin{eqnarray}
  0 = \Hgroup(\Xspace_0)  \to  \Hgroup(\Xspace_1)  \to  \ldots
         \to  \Hgroup(\Xspace_n) = \Hgroup(\Xspace,\Xspace^0)
         \to  \ldots \to  \Hgroup(\Xspace,\Xspace^n) = 0 .
  \label{eqn:extpersfilt}
\end{eqnarray}
We extend the notions of birth and death in the obvious way.
Since this filtration begins and ends with the zero group,
all classes eventually die. 
We also extend the graphical representation of the information contained
by forming \emph{persistence diagrams}, which we now introduce more formally.
We have such a diagram for each dimension $p$; see Figure \ref{fig:subdiagrams}.
Each diagram is a multiset of points in the plane,
containing one point $(r_i,r_j)$ for each coset of classes
that is born at $\Xspace_i$ or $(\Xspace,\Xspace^{n-i+1})$, and dies
entering $\Xspace_j$ or $(\Xspace,\Xspace^{n-j+1})$.
In some circumstances, it is convenient to add the points on the diagonal
to the diagram, but in this paper, we will refrain from doing so.
The persistence diagram contains three important subdiagrams,
corresponding to three different combinations of birth and death location.
The \emph{ordinary subdiagram}, $\Odgm{p}{f}$, represents classes
that are born and die during the first half of (\ref{eqn:extpersfilt}).
The \emph{relative subdiagram}, $\Rdgm{p}{f}$, represents classes
that are born and die during the second half.
Finally, the \emph{extended subdiagram}, $\Edgm{p}{f}$, represents classes
that are born during the first half and die during the second half
of the extended filtration.
Note that points in $\Odgm{p}{f}$ all lie above the main diagonal
while points in $\Rdgm{p}{f}$ all lie below.
On the other hand, $\Edgm{p}{f}$ may contain points on either side
of the main diagonal.
By $\Ddgm{}{f}$, we mean the points of all diagrams in all dimensions.
Drawing these subdiagrams side by side can be cumbersome,
and drawing them on top of each other can be confusing.
In Section \ref{sec3}, we will introduce a new design that addresses these concerns.

\subsection{Mixed Maps}
\label{sec22}

We note that the homology groups in the extended filtration
of (\ref{eqn:extpersfilt}), or in the shorter filtration of (\ref{eqn:persfilt}),
are all vector spaces over a fixed field and that the maps between
them are all linear maps.
In \cite{CadS10}, Carlsson and de Silva generalize this situation
to sequences of vector spaces that are connected by maps going
from left to right or from right to left.
We now briefly review their work as well as the related work
on level set zigzag modules in \cite{CSM09}.

\paragraph{Zigzag modules.}
A \emph{zigzag module} $\Wgroup$ is a finite sequence of vector spaces
connected by linear maps which either go forward or backward
between consecutive spaces:
\begin{equation}
  \Wgroup_1 \tot \Wgroup_2 \tot \ldots \tot \Wgroup_j \tot \Wgroup_{j+1}
            \tot \ldots \tot \Wgroup_n .
  \label{zz:module}
\end{equation}
If the arrow advances from $\Wgroup_j$ to $\Wgroup_{j+1}$,
then we denote the corresponding linear map as $a_j: \Wgroup_j \to \Wgroup_{j+1}$;
otherwise, we write $b_j: \Wgroup_{j+1} \to \Wgroup_j$.
A \emph{submodule} $\Ugroup$ of $\Wgroup$ is a collection of
linear subspaces $\Ugroup_j \subseteq \Wgroup_j$ such that
$a_j(\Ugroup_j) \subseteq \Ugroup_{j+1}$ or $b_j(\Ugroup_{j+1}) \subseteq \Ugroup_j$,
whichever is the case for $j$.
A submodule $\Ugroup$ is a \emph{summand} if there is a complementary submodule
$\Vgroup$, meaning every vector space splits as a direct sum 
$\Wgroup_j = \Ugroup_j \oplus \Vgroup_j$.
The authors in \cite{CadS10} prove that every zigzag module can be split into
indecomposable summands of a certain form, and, in particular, it has a basis,
a concept we now describe.
First, we suppose that we have, for each $j$,
a set of elements $u_j^i \in \Wgroup_j$ such that the nonzero elements
form a basis of $\Wgroup_j$.
In other words, we can decompose $\Wgroup_j$ into the direct sum
$\Wgroup_j = \bigoplus_i \spans{u_j^i}$,
noting that some of the terms on the right hand side may be zero.
We use the superscripts to form correspondences between the bases.
Specifically, we require $a_j(u_j^i) = u_{j+1}^i$, or $b_j(u_{j+1}^i) = u_j^i$,
depending on the case.
Furthermore, we assume that, for each superscript $i$, there exist $x \leq y$ such
that $u_j^i \neq 0$ iff $j \in [x,y]$.
In other words, for each fixed $i$, we have a submodule
\begin{equation}
  \spans{u_1^i} \tot \spans{u_2^i} \tot \ldots \tot \spans{u_j^i}
                \tot \spans{u_{j+1}^i} \tot \ldots \tot \spans{u_n^i}
  \label{zz:interval}
\end{equation}
of $\Wgroup$ in which the non-zero vector spaces are $1$-dimensional
and form a single contiguous subsequence connected by identity maps.
Calling such a submodule an \emph{interval module},
we think of it as being in correspondence with the closed interval $[x,y]$.
The collection $\{u_j^i\}$ is a \emph{basis} for the zigzag module
if $\Wgroup$ can 
be decomposed into the direct sum of the interval modules (\ref{zz:interval}).
Equivalently, the collection is a basis for $\Wgroup$ if each
map $a_j$ is the direct sum of the maps $\spans{u_j^i} \to \spans{u_{j+1}^i}$,
and each map $b_j$ is the direct sum of the maps
$\spans{u_{j+1}^i} \to \spans{u_j^i}$, whichever one is defined.

Although a zigzag module $\Wgroup$ can have many different bases, the set
of intervals associated to any such basis will be unique \cite{CadS10}.
For example, any basis for the zigzag module given by
the filtration in (\ref{eqn:persfilt}) will have one interval $[x,y]$
for each coset of classes born at $\Xspace_x$ and dying entering $\Xspace_y$.

\paragraph{Mayer-Vietoris diamonds.}

We are interested in an elementary operation that connects two minimally
different zigzag modules: a \emph{Mayer-Vietoris diamond}.
We suppose that we have two zigzag modules differing only at position $j$, and that at this position we have a diamond of the following form:
\begin{equation}
    \begin{diagram}
        \node[2]{\Hgroup(\Vspace,\Vspace')}                                   \\
        \node   {\Tot \Hgroup(\Cspace,\Cspace')}    \arrow{ne,t}{a_{j-1}}                     
        \node[2]{\Hgroup(\Dspace,\Dspace') \Tot}    \arrow{nw,t}{b_j}         \\
        \node[2]{\Hgroup(\Espace,\Espace')}         \arrow{ne,t}{a_j}\arrow{nw,t}{b_{j-1}}
    \end{diagram},
    \label{zz:diamond}
\end{equation}
where we show the more general, relative form in which
the primed spaces are subspaces of the corresponding unprimed ones,
and we have $\Espace = \Cspace \capsp \Dspace$, $\Espace' = \Cspace' \capsp \Dspace'$,
$\Vspace = \Cspace \cupsp \Dspace$, and $\Vspace' = \Cspace' \cupsp \Dspace'$.
We get the more special, absolute form by setting
$\Cspace' = \Dspace' = \Espace' = \Vspace' = \emptyset$.
The name of the diamond is justified by the long exact sequence we get by reading
the diamond from bottom to top and iterating through the dimensions.
When the primed spaces are all empty, this gives the classic version
of the Mayer-Vietoris sequence, and more generally,
we get the relative version:
$$
  \ldots \to \Hgroup_p(\Espace,\Espace')
         \to \Hgroup_p(\Cspace,\Cspace') \oplus \Hgroup_p(\Dspace,\Dspace')
         \to \Hgroup_p(\Vspace,\Vspace')
         \to \Hgroup_{p-1}(\Espace,\Espace') \to \ldots ;
$$
see e.g.\ \cite{Mun84}.
Importantly, this sequence is exact, which means that the image of each map
equals the kernel of the next map.

Such diamonds arise in the following context.
Consider again the function $f: \Xspace \to \Rspace$ and the interleaved sequence
of homological regular and critical values:
$s_0 < r_1 < s_1 < \ldots < r_n < s_n$.
Setting $\Wgroup_{2j} = \Hgroup (f^{-1} (s_j))$ and
$\Wgroup_{2j+1} = \Hgroup (f^{-1} [s_j, s_{j+1}])$,
we get a zigzag module of length $2n+1$,
which, following \cite{CSM09}, we refer to as the \emph{level set zigzag} of $f$.
It starts and ends with $0$ and alternates between advancing maps $a_{2j}$ and backward maps $b_{2j+1}$.
From this module, we can create a new one by fixing an index $j$, substituting
$[s_j, s_{j+2}] = [s_j, s_{j+1}] \cupsp [s_{j+1}, s_{j+2}]$ for
$s_{j+1} = [s_j, s_{j+1}] \capsp [s_{j+1}, s_{j+2}]$, and leaving all other groups unchanged;
of course we also reverse the two maps involving the changed space.
This produces a new zigzag module which differs from the old via a Mayer-Vietoris diamond.
This construction can be generalized by flipping between intersections and unions of larger intervals
and pairs of intervals, thus producing
a whole array of zigzag modules which differ via Mayer-Vietoris diamonds.

\begin{figure}[hbt]
 \vspace*{0.1in}
 \centering
 \resizebox{!}{3.0in}{\input{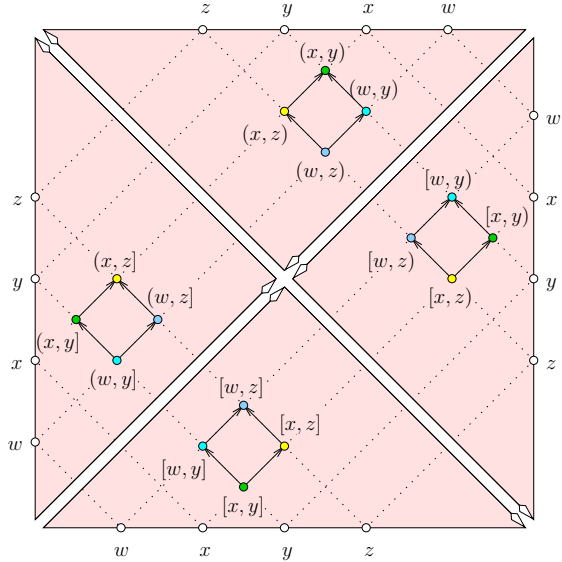}}
 \caption{Points in the pyramid are absolute and relative homology groups.
   Monotonic paths are zigzag modules,
   any two of which differ by a finite number of Mayer-Vietoris diamonds.}
 \label{fig:pyramid-1}
\end{figure}

\paragraph{The pyramid.}
Starting with the level set zigzag, we get an array of zigzag modules
which are best described as monotonic paths that go diagonally up and down,
always from left to right.
The array of such paths is connected within a pyramidal structure,
which we now describe.
As a graphical guide, we consider the square drawn in Figure \ref{fig:pyramid-1}.
We give it a coordinate system by parameterizing the downward slope
from $\infty$ at the upper left, to $-\infty$ in the middle,
and back up to $\infty$ at the lower right.
Similarly, we parameterize the upward slope from $-\infty$ at the lower left,
to $\infty$ in the middle, and back to $-\infty$ at the upper right.
The two slopes divide the square into four triangular regions,
each containing a point with coordinates $a$ and $b$ for every
choice of $a \leq b$.
We interpret this point differently in each of the regions.
To explain this interpretation, it is convenient to introduce a shorthand
that uses open set notation for pairs of closed sets,
writing $\Aspace - \Aspace'$ for $(\Aspace, \Aspace')$.
Specifically,
\begin{eqnarray*}
  f^{-1} (x,y]  &=&  ( f^{-1}(-\infty,y], f^{-1}(-\infty,x] ) , \\
  f^{-1} [x,y)  &=&  ( f^{-1}[x, \infty), f^{-1}[y, \infty) ) , \\
  f^{-1} (x,y)  &=&  ( f^{-1}(-\infty,\infty), f^{-1}(-\infty,x] \cupsp f^{-1} [y,\infty) ) .
\end{eqnarray*}
If a point with coordinates $x$ and $y$ lies in the bottom region,
we think of it as the space $f^{-1} [x,y]$.
However, if the point lies in the left, right, or top region,
we think of it as $f^{-1}(x,y]$, $f^{-1}[x,y)$, or $f^{-1}(x,y)$, respectively.
If we now take $w < x < y < z$ and consider the points
$(w,y)$, $(w,z)$, $(x,y)$, and $(x,z)$, we get a Mayer-Vietoris diamond
in each region; see Figure \ref{fig:pyramid-1}.
This is easiest to see in the closed interval case since
$[x,y] = [w,y] \capsp [x,z]$ and $[w,z] = [w,y] \cupsp [x,z]$.
In the closed-open case, we have
$[x,\infty) = [w,\infty) \capsp [x,\infty)$ and
$[w,\infty) = [w,\infty) \cupsp [x,\infty)$ as well as
$[z,\infty) = [z,\infty) \capsp [y,\infty)$ and
$[y,\infty) = [z,\infty) \cupsp [y,\infty)$.
Similar computations verify the diamond in the remaining two cases.

By repeated application of the diamond, we can generate any monotonic path
from the one along the bottom edge of the square.
Each path is thus decorated by spaces as described,
and applying the homology functor gives a zigzag module of
absolute and relative homology groups.
The latter arise when we move the left or right end of the path,
which can be done without the Mayer-Vietoris diamond because the corresponding
spaces are and stay empty so that the module remains unchanged.
Besides the level set zigzag along the bottom edge,
we are particularly interested in the path along the upward slope,
which translates into the extended filtration of (\ref{eqn:extpersfilt}).
Its midpoint is $(-\infty,\infty)$, the center of the square,
which results in $\Hgroup(f^{-1}(-\infty,\infty)) = \Hgroup(\Xspace)$.
For this reason, we think of the center as the apex of a pyramid,
as viewed from above.

\begin{result}[Remark]
  As a partial justification for the notation with open sets,
  we mention that the homology group of the preimage of
  the interval $(x,y)$, if computed with infinite chains,
  is isomorphic to the relative homology group of  
  $(f^{-1} [x,y], f^{-1} (x) \cupsp f^{-1} (y))$. By excision, this
  is isomorphic to the relative homology group of 
  $(f^{-1}(-\infty,\infty], f^{-1} (-\infty, x] \cupsp f^{-1} [y, \infty))$.
\end{result}

\subsection{Perturbations}
\label{sec23}

The reader who wishes to learn how to read the homology of interlevel sets
can safely skip Section \ref{sec23} and now continue with Section \ref{sec3}.
However, to differentiate the robust from the non-robust homological information
in these readings, we need to first understand the subgroups of homology
that give meaning to this concept.

\paragraph{Well groups.}
Suppose that we have a continuous mapping $f: \Xspace \to \Yspace$ between topological spaces.
Given a subset $\Aspace \subseteq \Yspace$, we review here
the definition of the well groups $\Ugroup_{\Aspace}(r)$
for each radius $r \geq 0$.
When $\Aspace$ is clear from context,
we will drop it from the notation and simply write $\Ugroup(r)$,
by which we mean the direct sum of groups $\Ugroup_p (r)$,
for each homological dimension $p$.
We will also need the assumption that $f^{-1}(\Aspace)$
has homology groups of finite rank in each dimension.
In addition to the mapping $f$, we assume a subspace $\Pcal$ of $C(\Xspace, \Yspace),$ the space
of continuous mappings from $\Xspace$ to $\Yspace$, requiring that $\Pcal$ contains $f$.
For example, $\Pcal$ might consist of all mappings homotopic to $f$.
We assume a metric on $\Pcal$ and write $\Pdist{f}{h}$ for the distance between two mappings.
We call $h$ an \emph{$r$-perturbation} of $f$ if $\Pdist{f}{h} \leq r$.
Given $\Aspace \subseteq \Yspace$, we introduce the \emph{radius function},
$f_{\Aspace} : \Xspace \to \Rspace$, by setting $f_{\Aspace}(x)$ to
the infimum value of $r$ for which there exists an $r$-perturbation $h \in \Pcal$ with
$h(x) \in \Aspace$.
We filter $\Xspace$ via the sublevel sets of the radius function, setting
$\Xspace_r(f_{\Aspace}) = f_{\Aspace}^{-1}[0,r]$.
For $r < s,$ there is a map
$\fmap_\Aspace^{r,s}: \Hgroup(\Xspace_r(f_{\Aspace})) \to \Hgroup(\Xspace_s(f_{\Aspace}))$.
The preimage of $\Aspace$ under any $r$-perturbation $h$ of $f$ will obviously
be a subset of $\Xspace_r(f_{\Aspace})$,
and hence there is a map on homology,
$\jmap_h: \Hgroup(h^{-1}(\Aspace)) \to \Hgroup(\Xspace_r(f_{\Aspace}))$.
Given a class $\alpha \in \Hgroup(\Xspace_r(f_{\Aspace}))$ and an $r$-perturbation $h$
of $f$, we say that $\alpha$ is \emph{supported} by $h$
if $\alpha \in \image{\jmap_h}$.
The \emph{well group} $\Ugroup(r) \subseteq \Hgroup(\Xspace_r(f_{\Aspace}))$ is then
defined \cite{EMP10} to consist of the classes that are supported
by all $r$-perturbations of $f$:
\begin{eqnarray*}
  \Ugroup(r)  &=&  \bigcap_{\Pdist{h}{f} \leq r} \image{\jmap_h} .
\end{eqnarray*}
For $r < s$, the map $\fmap_\Aspace^{r,s}$ restricts to a map
$\Ugroup(r) \to \Hgroup(\Xspace_s(f_{\Aspace}))$.
On the other hand,
$\Hgroup(\Xspace_s(f_{\Aspace}))$ contains $\Ugroup(s)$ as a subgroup.
It can be shown that $\Ugroup(s) \subseteq \fmap_\Aspace^{r,s}(\Ugroup(r))$
whenever $r < s$; see \cite{EMP10}.
In other words, the rank of the well group can only decrease
as the threshold value increases.
We call a value of $r$ at which the rank of the well group
decreases a \emph{terminal critical value} of $f_\Aspace$.
The \emph{well diagram} of $f$ and $\Aspace$ is the multiset
of terminal critical values of $f_\Aspace$,
taking a value $k$ times if the rank of the well group
drops by $k$ at the value.
Often we will refer to this diagram as the \emph{robustness} of
the preimage $f^{-1}(\Aspace)$.
In this paper, we focus on the case $\Yspace = \Rspace$ and $\Pcal = C(\Xspace,\Rspace)$,
lifting the usual metric on $\Rspace$
to $\Pcal$ by defining $\Pdist{f}{h} = \Maxdist{f}{h} = \sup_{x \in \Xspace} |f(x) - h(x)|$.
In this case, the radius function satisfies $f_{\Aspace}(x) = \inf_{a \in \Aspace} |f(x) - a|$.
In general, the relationship between the terminal critical values
and the homological critical values of $f_\Aspace$ is not
completely understood.
However, if $\Yspace = \Rspace$ and $\Aspace$ is a point, we will see that the former is a subset
of the latter.
We get more complicated relationships when $\Aspace$ is an interval.

\begin{figure}[hbt]
 \vspace*{0.1in}
 \centering
 \resizebox{!}{1.8in}{\input{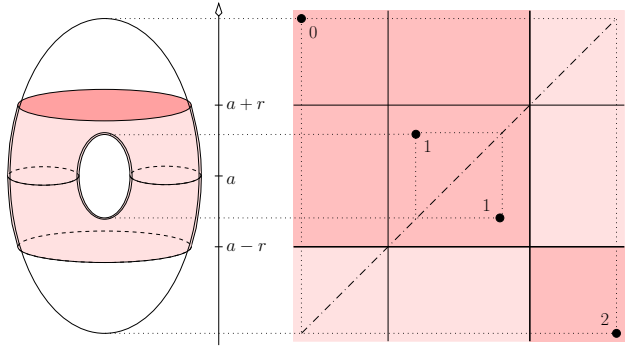}}
 \caption{Left: the torus and the preimage of the interval $[a-r,a+r]$.
          Right: the extended persistence diagram of the vertical height function.
          Each point is labeled by the dimension
          of the corresponding homology class.
          The dark shaded portions of the diagram represent the homology
          of $f^{-1} [a-r, a+r]$.}
\label{fig:twopants}
\end{figure}

\paragraph{Example.}
Consider the torus $\Xspace$,
as shown in Figure \ref{fig:twopants}, along with the vertical
height function $f: \Xspace \to \Rspace$ and the space $\Aspace = \{a\}$.
The preimage of $\Aspace$, $f^{-1} (\Aspace) = f_\Aspace^{-1}(0)$,
consists of two disjoint circles on the torus;
hence there are two components and two independent $1$-cycles,
all belonging to the well group at radius $0$.
For small values of $r$, $\Xspace_r (f_\Aspace)$
consists of two disjoint cylinders.
The homology has yet to change; furthermore, although the proof will
come later, all classes still belong to the well groups at these small radii.

Now consider the value of $r$ shown in Figure \ref{fig:twopants}.
For this $r$, the sublevel set $\Xspace_r = \Xspace_r(f_{\Aspace})$ consists of
two pair-of-pants glued together along two common circles.
We note that $\Hgroup_0(\Xspace_r)$ has dropped in rank by one,
while the rank of $\Hgroup_1 (\Xspace_r)$ has grown to three.
In contrast, the rank of $\Ugroup_1(r)$ is less than or equal to one.
Indeed, the function $h: \Xspace \to \Rspace$, defined by $h = f - r$,
is an $r$-perturbation of $f$ and the zero set
of the corresponding distance function, $h_\Aspace^{-1} (0) = f^{-1}(a+r)$,
is a single closed curve.
Since the rank of the first homology group of that curve is one,
and since the rank of $\image{\jmap_h}$ can be no bigger than this rank,
the well group $\Ugroup_1 (r)$ can also have rank at most one.
That it does in fact have rank exactly one will follow from our
results in Section \ref{sec4}.

\paragraph{Relative well groups.}
Since the pyramid involves relative homology groups, it seems wise
to extend the definition of well groups into the context of relative homology.
While this notion is new, it follows the above ideas closely so that
presenting the definition in this background section seems appropriate.
Assume again that we have a continuous mapping $f: \Xspace \to \Yspace$
between topological spaces, as well as a subspace $\Pcal$ of $C(\Xspace,\Yspace)$
that contains $f$ and is equipped with a metric. 
Given a nested pair $\Aspace' \subseteq \Aspace$ of subspaces
of $\Yspace$, and a radius $r \geq 0$, we note
that $\Xspace_r' = \Xspace_r(f_{\Aspace'})$ is a subset of
$\Xspace_r = \Xspace_r(f_{\Aspace})$.
For each $r$-perturbation $h$ of $f$, there is an inclusion of pairs
$(h^{-1}(\Aspace),h^{-1}(\Aspace')) \hookrightarrow (\Xspace_r, \Xspace_r'),$
which induces a map
$j_h: \Hgroup(h^{-1}(\Aspace),h^{-1}(\Aspace')) \to \Hgroup(\Xspace_r, \Xspace_r')$
between relative homology groups.
The \emph{relative well group} $\Ugroup_{(\Aspace,\Aspace')}(r)$ is defined
to be the intersection of the images of these maps, taken over all $r$-perturbations of $f$:
\begin{eqnarray*}
  \Ugroup_{(\Aspace,\Aspace')}(r)  &=&  \bigcap_{\Pdist{h}{f} \leq r} \image{j_h} .
\end{eqnarray*}
When a distinction is needed, we will refer to the previous notion of well groups
as \emph{absolute} well groups.

\section{Combinatorics of Homology}
\label{sec3}

In this section, we present the first half of our point calculus,
showing how to read the homology of a level or interlevel set from
the extended persistence diagram.
The crucial technical concept is that of a basis of the pyramid of zigzag modules,
which we establish by strengthening the Pyramid Theorem in \cite{CSM09}.

\paragraph{Flipping a basis.}
We construct a basis for the pyramid one step at a time,
by flipping the basis of one zigzag module to the next.
For this purpose, we consider two zigzag modules that differ
at one position, and we assume that there is a Mayer-Vietoris diamond
serving as a connecting bridge between the two modules at that position.
Drawing the diamond with the intersection at the bottom and the union at the top,
as in (\ref{zz:diamond}), we say the diamond connects the
\emph{lower} module with the \emph{upper} module.
Given a basis of the lower module, we can show that we can construct a basis of the
upper module so that the two bases agree on the overlap.
We refer to this operation as \emph{flipping} the first basis to the second.
\begin{result}[Basis Flip Lemma]
  Given two zigzag modules that differ by a single Mayer-Vietoris diamond,
  we can flip any basis of the lower module to a basis of the upper module.
\end{result}
\proof
 We give a proof by construction.
 Writing $\{e_k^i\}$ for the basis of the lower zigzag module,
 we describe a basis $\{v_k^i\}$ of the upper zigzag module
 that differs from the lower one only at the position $j$
 at which the modules differ; as in (\ref{zz:diamond}).
 We thus at once set $v_k^i = e_k^i$ for all $k \neq j$, and the main
 task is then the construction of the $v_j^i$.
 Put briefly, our rule will be that $v_j^i \neq 0$ iff an odd number
 of $e_{j-1}^i$, $e_j^i$, $e_{j+1}^i$ are non-zero.
 We give more specifics via a case analysis.
 The cases are labelled pictorially, with black dots
 denoting non-zero classes, showing only the positions $j-1,j,j+1$.
 \begin{description}\denselist
   \item[{\sc Case 1 (\raisebox{-.85ex}{\includegraphics{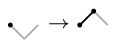}}):}]
     We have $e_{j-1}^i \neq 0$ and $e_j^i = e_{j+1}^i = 0$,
     and define $v_j^i$ as well as the advancing map using the Mayer-Vietoris diamond,
     namely $v_j^i = a_{j-1}(e_{j-1}^i)$, which is non-zero
     by exactness and because $e_j^i = 0$.
   \item[{\sc Case 2 (\raisebox{-.85ex}{\includegraphics{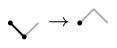}}):}]
     Again we set $v_j^i = a_{j-1}(e_{j-1}^i),$ which is zero
     by exactness and because $e_j^i \neq 0$.
   \item[{\sc Case 3 (\raisebox{-.85ex}{\includegraphics{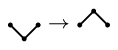}}):}]
     We set $v_j^i = a_{j-1}(e_{j-1}^i) = b_j(e_{j+1}^i)$,
     which in this case is non-zero.
     Indeed, if it were zero, then, by exactness, the pair $(e_{j-1}^i,0)$
     would be in the image of the map $b_{j-1} \oplus a_j$.
     By the direct-sum decomposition of the maps in the basis,
     this would imply that $a_j(e_j) = 0$, a contradiction.
   \item[{\sc Case 4 (\raisebox{-.85ex}{\includegraphics{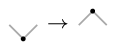}}):}]
     We have $e_j^i \neq 0$ and $e_{j-1}^i = e_{j+1}^i = 0$.
     If there are $\ell \geq 0$ indices $i$ of this kind, then the orthogonal complement
     to the image of the map $c_j$, defined below, has rank $\ell$, as we prove shortly.
     We pick $\ell$ classes $v_j^i$ that span this complement.
     Since $v_j^i$ maps to $e_j^i$ via the connecting homomorphism of the
     Mayer-Vietoris sequence, the homological dimension of $v_j^i$ is one
     higher than that of $e_j^i$.
   \item[{\sc Case 5 (\raisebox{-.85ex}{\includegraphics{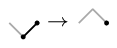}}):}]
     This is symmetric to Case 2, and we set $v_j^i = b_j(e_{j+1}^i) = 0$.
   \item[{\sc Case 6 (\raisebox{-.85ex}{\includegraphics{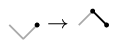}}):}]
     This is symmetric to Case 1, and we set $v_j^i = b_j(e_{j+1}^i) \neq 0$.
 \end{description}
 Note first that we now have interval modules $\{e_j^i\}$
 in the lower zigzag module, and interval modules $\{v_j^i\}$
 in the upper zigzag module.
 To show that the latter are indeed summands, we only need to verify that
 the non-zero classes $v_j^i$ form a basis of $\Hgroup(\Vspace, \Vspace')$,
 the new group in the upper zigzag module.
 Using the notation in (\ref{zz:diamond}), we let $\Egroup$ denote the vector space
 spanned by the pairs $(e_{j-1}^i,e_{j+1}^i)$, noting that $\Egroup$
 is a subspace of $\Hgroup(\Cspace, \Cspace') \oplus \Hgroup(\Dspace, \Dspace')$,
 but because of Case 3 it is not necessarily the entire direct sum.
 We consider the subspaces $\Egroup_N$ of
 $\Egroup$ spanned
 by the pairs $(e_{j-1}^i,e_{j+1}^i)$ in each Case $N$, for $1 \leq N \leq 6$.
 These subspaces are independent and span the entire space $\Egroup$.
 In other words, zero is the only element common to any two of the subspaces,
 and the ranks of the subspaces add up to the rank of $\Egroup$.

 The case analysis suggests a map 
 $c_j: \Egroup   \to    \Hgroup(\Vspace, \Vspace')$
 with $c_j((e_{j-1}^i, e_{j+1}^i)) = v_j^i$, if $(e_{j-1}^i, e_{j+1}^i) \neq (0,0)$, and zero otherwise.
 Since $\Egroup_4 = 0$, this map is zero on $\Egroup_4$, but it is also
 zero on $\Egroup_2$ and $\Egroup_5$.
 Furthermore, $c_j$ is injective when restricted to $\Egroup_1$, $\Egroup_3$, and $\Egroup_6$.
 We proceed to show that the images of these latter three vector spaces under $c_j$ are
 independent of one another.
 To derive a contradiction, we first suppose that $c_j(\Egroup_1) \capsp c_j(\Egroup_6)$
 contains a non-zero class.
 Then there must exist
 $(\alpha,0) \in \Egroup_1$ and $(0,\beta) \in \Egroup_6$ with $a_{j-1}(\alpha) = b_j(\beta) \neq 0$.
 Hence, $(\alpha,\beta) \in \kernel{(a_{j-1} \oplus b_j)}$,
 which, by exactness, tells us that $\alpha \in \image{b_{j-1}}$.
 But this contradicts the direct-sum decomposition of the map $b_{j-1}$.
 Next, suppose that $c_j(\Egroup_1) \capsp c_j(\Egroup_3)$ contains a non-zero class, which means
 there exists $(\alpha,0) \in \Egroup_1$ and $(\gamma,\beta) \in \Egroup_3$
 such that $a_{j-1}(\alpha) = b_j(\beta) \neq 0$.
 As above, this implies that $(\alpha,\beta) \in \kernel{(a_{j-1} \oplus b_j)}$,
 and we reach the same contradiction.
 Finally, a symmetric argument gives $c_j(\Egroup_3) \capsp c_j(\Egroup_6) = 0$.
 We conclude that $c_j(\Egroup_1)$, $c_j(\Egroup_3)$, and $c_j(\Egroup_6)$ are independent subspaces
 of $\Hgroup(\Vspace, \Vspace')$.
 In Case 4, we picked a basis for the orthogonal complement to their span;
 all together, we have a basis of $\Hgroup(\Vspace, \Vspace')$, as required.
\eop

\begin{figure}[hbt]
 \vspace*{0.1in}
 \centering
 \resizebox{!}{3.6in}{\input{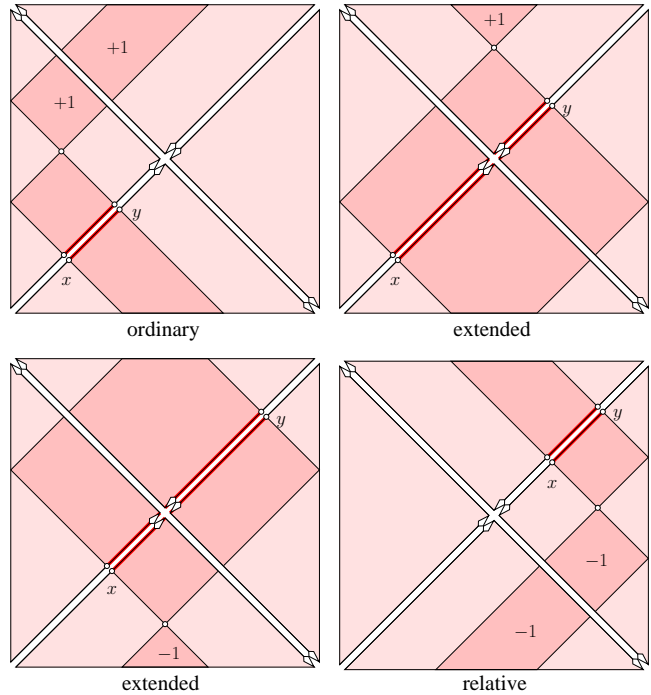}}
    \caption{The basis element that corresponds to the interval from $x$ to $y$
        along the upward slope maps to all spaces between the paths of its two endpoints.
        The four squares show the pattern for the four different types of intervals.}
    \label{fig:transformation}            
\end{figure}

\paragraph{Establishing a basis.}
The Pyramid Theorem in \cite{CSM09} establishes
an explicit bijection between the interval modules that arise in the decomposition
of any two zigzags within the pyramid.
We strengthen this result by establishing bases on all the zigzag modules in such a way that
the basis elements correspond to the intervals and respect the same bijections.
We call this a \emph{basis} of the pyramid.
To construct it, we note that the paths in the pyramid are connected by Mayer-Vietoris diamonds.
We can therefore flip a basis of the level set zigzag upwards through
the entire pyramid via repeated application of the Basis Flip Lemma.
\begin{result}[Pyramid Basis Theorem]
  A basis of the level set zigzag module extends to a basis of the entire pyramid.
\end{result}
We now give an explicit description of how the interval modules of the various paths
in the pyramid relate to each other.
A convenient reference in this description
is the extended filtration (\ref{eqn:extpersfilt}),
which follows the upward slope through the middle of the pyramid.
Its first half is parameterized from $- \infty$ to $\infty$,
and its second half from $\infty$ back to $- \infty$.
Let now $x$ and $y$ be two points along the upward slope,
with $x$ to the left of $y$.
We distinguish between the ordinary case ($x < y$, both in the first half),
the relative case ($y < x$, both in the second half),
and the two extended cases ($x < y$ and $y < x$, with $x$ in the first half
and $y$ in the second half).
For each case, we sketch how the basis element of the interval
corresponds to basis elements of other homology groups in Figure \ref{fig:transformation}.
As a general pattern, the two points trace out two curves
consisting of segments with slopes $\pm 45^\circ$ that reflect before they
hit the vertical sides and end at the horizontal sides of the square.
The reason for the slopes are Cases 1, 2, 5, and 6 in the proof of the
Basis Flip Lemma, and the reason for the reflection is the local change
in the zigzag structure caused by moving the terminal zero group up.
The two curves cross at one point inside the square, and the location of that
point is characteristic for the case
(the triangular region on the left in the ordinary case,
at the top and at the bottom in the two extended cases,
and on the right in the relative case).
The crossing is caused by Case 4,
in which the correspondence between the basis elements is constructed
via the connecting homomorphism of the Mayer-Vietoris sequence
and therefore comes with a shift in homological dimension.

\paragraph{Turning the table.}
The regions in Figure \ref{fig:transformation} show all the spaces
represented by points in the pyramid to which the basis element
corresponding to the interval $[x,y]$ is relevant.
We are now interested in the inverse question:
which basis elements are relevant to a given space?
More specifically: which intervals in the decomposition
of the extended filtration (\ref{eqn:extpersfilt}) map to the basis
of the homology group of the space represented by a point with coordinates $a$ and $b$?
We answer this question by considering the following subregions
of the $p$-dimensional persistence diagram:
\begin{eqnarray*}
  \lRegion{p}[a,b]  &=&  \{ (x,y) \in \Odgm{p}{f} \mid x < b < y \}
                     \sqcup  \{ (x,y) \in \Edgm{p}{f} \mid x < b, a < y \} , \\
  \rRegion{p}[a,b]  &=&  \{ (x,y) \in \Edgm{p}{f} \mid b < x, y < a \} 
                     \sqcup  \{ (x,y) \in \Rdgm{p}{f} \mid y < a < x \}, \\
  \lRegion{p}[a,b)  &=&  \{ (x,y) \in \Edgm{p}{f} \mid a < y < b \}
                     \sqcup  \{ (x,y) \in \Rdgm{p}{f} \mid a < y < b < x \} , \\
  \rRegion{p}[a,b)  &=&  \{ (x,y) \in \Rdgm{p}{f} \mid y < a < x < b\}, \\
  \lRegion{p}(a,b]  &=&  \{ (x,y) \in \Odgm{p}{f} \mid x < a < y < b \}, \\
  \rRegion{p}(a,b]  &=&  \{ (x,y) \in \Odgm{p}{f} \mid a < x < b < y\}
                     \sqcup  \{ (x,y) \in \Edgm{p}{f} \mid a < x < b \}, \\
  \lRegion{p}(a,b)  &=&  \{ (x,y) \in \Odgm{p}{f} \mid x < a < y \} 
                     \sqcup  \{ (x,y) \in \Edgm{p}{f} \mid x < a, b < y \} , \\
  \rRegion{p}(a,b)  &=&  \{ (x,y) \in \Edgm{p}{f} \mid a < x, y < b \}
                     \sqcup  \{ (x,y) \in \Rdgm{p}{f} \mid y < b < x \},
\end{eqnarray*}
where we assume that $a$ and $b$ are both homological regular values.
\begin{figure}[hbt]
 \vspace*{0.1in}
 \centering
 \resizebox{!}{2.8in}{\input{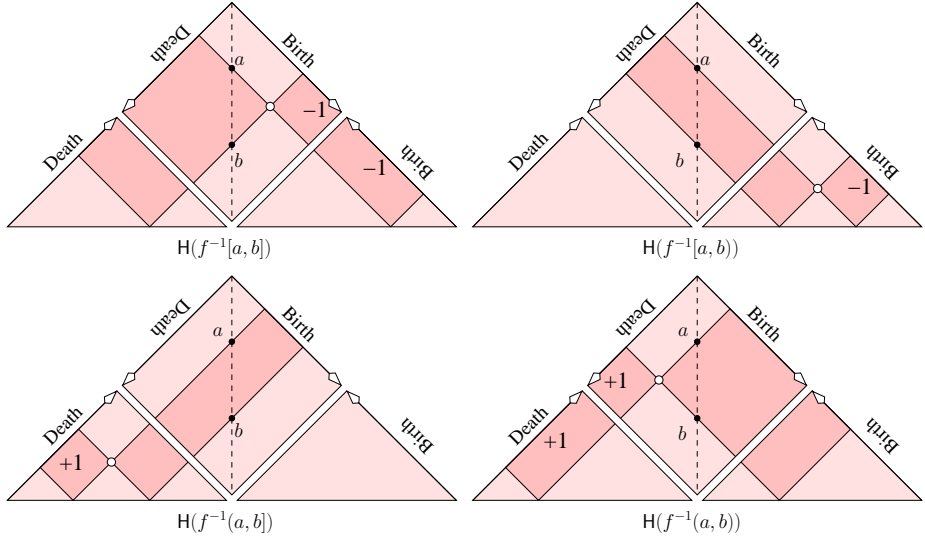}}
 \caption{The triangle design of the persistence diagram showing the regions
   $\lRegion{}$ and $\rRegion{}$ for the four types of intervals in darker shading.
   When we collect the points to compute the rank of the $p$-th homology group,
   we shift the homological dimension of classes as shown.}
 \label{fig:four-homology}
\end{figure}
These multisets are displayed in Figure \ref{fig:four-homology}
in which we have also introduced a new, and for our purposes more convenient,
way of drawing the extended persistence diagram.
We have glued the domains of the three sub-diagrams and drawn
the result as a right-angled triangle.
In this triangle, the birth and death axes go from $- \infty$
up to $+ \infty$ and then continue on back to $- \infty$.
In other words, we flip the extended subdiagram upside down
and glue its (formerly) upper side to the upper side of the
ordinary subdiagram.
Similarly, we rotate the relative subdiagram by $180$ degrees
and glue its (formerly) right side to the right side of the
(flipped) extended subdiagram.
After gluing the three domains, we rotate the design by
$-45$ degrees so the triangle rests on its longest side,
consisting of the diagonals in the ordinary and relative subdiagrams.
The diagonal of the extended subdiagram is now the vertical symmetry
axis passing through the middle of the triangle.

\begin{result}[Remark]
  There is a straightforward translation of this triangular design
  to the representation of persistence advocated in \cite{CCGZ05}.
  Namely, draw a isosceles right-angled triangle downward from
  each point in the multiset and call the horizontal lower edge the corresponding \emph{bar}.
  The \emph{barcode} is the multiset of bars, one for each point in the diagram.
  Similarly, we can translate the triangular design into the square design
  of the pyramid by cutting along the vertical axis,
  turning the right triangle upside-down, and gluing the two triangles
  along their hypotenuses.
\end{result}

\paragraph{Reading interlevel sets.}
The purpose of the multisets defined above is to offer a convenient way to read
the absolute or relative homology of an interlevel set from the extended persistence diagram.
We need some definitions to combine all four types into one.
First, we let $\Bcal$ be the collection of interval modules in the decomposition
of the extended filtration (\ref{eqn:extpersfilt}). As mentioned earlier, this collection
is in bijective correspondence with the points in $\Ddgm{}{f}$.
We write $\Vgroup = \spans{\Bcal}$ for the abstract vector space spanned by $\Bcal$, and
we let $\Vcal = \{ \spans{\Bcal'} \mid \Bcal' \subseteq \Bcal\}$ be the collection
of vector spaces spanned by subsets of this basis.
Second, we write
\begin{eqnarray*}
  \Region{p} (I)  &=&  \left\{ \begin{array}{ll}
    \lRegion{p} [a,b]   \sqcup \rRegion{p+1} [a,b]  &  \mbox{\rm if~} I = [a,b] , \\
    \lRegion{p} [a,b)   \sqcup \rRegion{p+1} [a,b)  &  \mbox{\rm if~} I = [a,b) , \\
    \lRegion{p-1} (a,b] \sqcup \rRegion{p} (a,b]  &  \mbox{\rm if~} I = (a,b] , \\
    \lRegion{p-1} (a,b) \sqcup \rRegion{p} (a,b)  &  \mbox{\rm if~} I = (a,b) ,
  \end{array} \right.
\end{eqnarray*}
for the region of points in the persistence diagram that correspond to the basis elements
of $\Hgroup_p (f^{-1} (I))$, and call it a \emph{pair of wings}.
With these concepts, we have the following result,
which implies that the rank of $\Hgroup_p (f^{-1} (I))$
is the number of points in $\Region{p} (I)$:
\begin{result}[Interlevel Set Lemma]
  For each dimension $p$ and each interval $I$ whose endpoints are homological regular values,
  there exists an isomorphism that takes $\Hgroup_p (f^{-1} (I))$ to the vector space
  $\Ggroup_p (I) \in \Vcal$ spanned by the basis vectors corresponding to the points in
  $\Region{p}(I)$.
\end{result}
\proof
 Write $\Bcal = \{e^i\}$ and let $\{v^i\}$ be the basis of the group $\Hgroup_p(f^{-1} (I))$,
 where $I$ is an interval with endpoints $a \leq b$ that can be closed,
 closed-open, open-closed, or open.
 The claimed isomorphism is then the linear map $\gamma: \Hgroup_p(f^{-1}(I)) \to \Vgroup$
 defined by $\gamma(v^i) = \{e^i\}$ for all non-zero $v^i$.

 To understand why the image of $\gamma$ consists of the intervals
 that correspond to the points in $\Region{p} (I)$,
 we need to recall the transformation rules sketched in Figure \ref{fig:transformation}.
 Consider for example the closed interval case, $I = [a,b]$,
 for which $\Region{p} (I) = \lRegion{p} [a,b] \sqcup \rRegion{p+1} [a,b]$.
 Since the interval is closed, the homology group is represented by
 the point $(a,b)$ in the lower triangular region.
 To lie in the dark shaded region, this point must satisfy the constraint
 $x < b < y$ in the ordinary case,
 $x < b$ and $a < y$ in the first extended case,
 and $x < b$ and $a < y$ without dimension shift in the second extended case.
 These inequalities define $\lRegion{p} [a,b]$.
 Furthermore, we get $b < x$ and $y < a$ with dimension shift in the second
 extended case, and $y < a < x$, again with dimension shift, in the relative case.
 These inequalities define $\rRegion{p+1} [a,b]$, which completes the proof
 in the closed case.
 For a proof of the closed-open, open-closed, and open cases,
 note that the points representing
 $\Hgroup_p (f^{-1} (I))$ are found in the right, left, and top triangular region of the pyramid,
 and then argue in a similar fashion.
\eop

\section{Combinatorics of Robustness}
\label{sec4}

The definition of well group given in Section \ref{sec2} involves
an uncountable number of perturbations,
which give rise to the intersection of a potentially large number of subgroups,
and as such does not seem amenable to computation.
In this section, we show that the situation in the real-valued case is simpler,
and that we are able to read the absolute and relative well groups
directly from the extended persistence diagram.
We begin with a consequence of the Mayer-Vietoris sequence,
which provides the main technical ingredient of our proofs.

\paragraph{A corollary of Mayer-Vietoris.}
For convenience, we establish the following notational convention,
wherein we reuse the same letter in different fonts.
We will need it for absolute and for relative homology groups.
To avoid repetition, we state it now for the more general relative case.
Letting $\Uspace' \subseteq \Uspace$ and $\Vspace' \subseteq \Vspace$
be pairs of topological spaces,
we write $(\Uspace, \Uspace') \hookrightarrow (\Vspace, \Vspace')$
if $\Uspace \subseteq \Vspace$ and $\Uspace' \subseteq \Vspace'$.
This inclusion of pairs induces a map
$\umap: \Hgroup(\Uspace, \Uspace') \to \Hgroup(\Vspace, \Vspace')$
on homology groups, and we write $\Ugroup = \image{\umap}$ for the image of this map.
Note that $\Ugroup$ is always a subgroup of $\Hgroup(\Vspace, \Vspace')$,
namely the subgroup of homology classes that have a chain representative
carried by $(\Uspace, \Uspace')$.
Note also that the rank of $\Ugroup$ can never exceed
the rank of $\Hgroup(\Vspace, \Vspace')$.
Suppose that, furthermore, $(\Tspace, \Tspace') \hookrightarrow (\Uspace, \Uspace')$.
Then, from the sequence of maps
$\Hgroup(\Tspace, \Tspace') \to \Hgroup(\Uspace, \Uspace') \to \Hgroup(\Vspace, \Vspace')$,
we see that $\Tgroup$, the image of $\Hgroup(\Tspace, \Tspace')$ in
$\Hgroup(\Vspace, \Vspace')$, must be a subgroup of $\Ugroup$.
The following lemma is a direct consequence of the exactness of
the Mayer-Vietoris sequence.
However, we will use it often enough that it seems reasonable
to state and prove it formally.
\begin{result}[Mayer-Vietoris Lemma]
  Suppose the pair of topological spaces $\Vspace' \subseteq \Vspace$
  can be decomposed as $\Vspace = \Cspace \cupsp \Dspace$
  and $\Vspace' = \Cspace' \cupsp \Dspace'$,
  where $\Cspace' \subseteq \Cspace$ and $\Dspace' \subseteq \Dspace$.
  Set $(\Espace, \Espace') = (\Cspace \capsp \Dspace, \Cspace' \capsp \Dspace')$.
  If a class $\alpha \in \Hgroup(\Vspace, \Vspace')$ belongs to $\Cgroup$
  and to $\Dgroup$, then $\alpha$ also belongs to $\Egroup$.
\end{result}

\proof
 Following our convention, we use the notation
 $\cmap: \Hgroup(\Cspace, \Cspace') \to \Hgroup(\Vspace, \Vspace')$
 for the map on homology induced by the inclusion of $(\Cspace, \Cspace')$
 in $(\Vspace, \Vspace')$.
 Similarly, we write $\dmap: \Hgroup(\Dspace, \Dspace') \to \Hgroup(\Vspace, \Vspace')$
 and $\emap: \Hgroup(\Espace, \Espace') \to \Hgroup(\Vspace, \Vspace')$, 
 as well as $\emap_c: \Hgroup(\Espace, \Espace') \to \Hgroup(\Cspace, \Cspace')$
 and $\emap_d: \Hgroup(\Espace, \Espace') \to \Hgroup(\Dspace, \Dspace')$.
 Note that $\Cgroup = \image{\cmap}$, $\Dgroup = \image{\dmap}$,
 and $\Egroup = \image{\emap}$.
 Consider now the relevant portion of the Mayer-Vietoris sequence
 for $(\Vspace, \Vspace')$:
 \[ \begin{diagram}
   \node{\Hgroup(\Espace, \Espace')} \arrow{e,t}{(\emap_c,\emap_d)}
     \node{\Hgroup(\Cspace, \Cspace') \oplus \Hgroup(\Dspace, \Dspace')}
     \arrow{e,t}{\cmap - \dmap} \node{\Hgroup(\Vspace, \Vspace').}
 \end{diagram} \]
 By assumption, $\alpha \in \Cgroup$, so there exists some
 $\alpha_c \in \Hgroup(\Cspace, \Cspace')$ such that $\cmap(\alpha_c) = \alpha$.
 Similarly, there exists an $\alpha_d \in \Hgroup(\Dspace, \Dspace')$
 such that $\dmap(\alpha_d) = \alpha$. 
 This implies that the pair $(\alpha_c,\alpha_d)$
 belongs to the kernel of $\cmap - \dmap$, and thus also,
 by exactness of the sequence,
 belongs to the image of $(\emap_c,\emap_d)$.
 Hence, there exists $\alpha_e \in \Hgroup(\Espace, \Espace')$
 with $\emap_c(\alpha_e) = \alpha_c$ and $\emap_d(\alpha_e) = \alpha_d$.
 In particular, since $\emap = \cmap \circ \emap_c$,
 we have $\emap(\alpha_e) = \alpha$,
 and therefore $\alpha \in \Egroup$ as claimed.
\eop

In the typical application of the Mayer-Vietoris Lemma,
we will construct further pairs $(\Tspace, \Tspace') \hookrightarrow (\Cspace, \Cspace')$
and $(\Bspace, \Bspace') \hookrightarrow (\Dspace, \Dspace')$
such that $\alpha \in \Tgroup \capsp \Bgroup$.
From the remark above, we know that $\Tgroup \subseteq \Cgroup$
and $\Bgroup \subseteq \Dgroup$.
The lemma then applies and we can conclude that $\alpha \in \Egroup$, as before.

\paragraph{The well group of a level set.}
As a warm-up exercise, we first consider the case in which
$\Aspace$ is a single point.
More specifically, we suppose that we have a compact topological space $\Xspace$
and a function $f: \Xspace \to \Rspace$, and we find the well groups
$\Ugroup(r) = \Ugroup_{\Aspace}(r)$, where $\Aspace = \{a\}$ is some point on the real line.
In this case, $\Xspace_r(f_{\Aspace}) = f_{\Aspace}^{-1} [0,r]  = f^{-1} [a-r,a+r] $.
To state the formula, we distinguish two particular subspaces of
$\Xspace_r = \Xspace_r(f_{\Aspace})$,
namely the \emph{top level set}, $\Tspace_r = f^{-1}(a+r)$,
and the \emph{bottom level set}, $\Bspace_r = f^{-1}(a-r)$. 
Using the convention from before, we write $\Tgrouptoo_r$ and $\Bgroup_r$
for the images of $\Hgroup(\Tspace_r)$ and $\Hgroup(\Bspace_r)$
in $\Hgroup(\Xspace_r)$.
\begin{result}[Point Formula]
  $\Ugroup(r) = \Tgrouptoo_r \capsp \Bgroup_r$, for every $r \geq 0$.
\end{result}
\proof
 We prove equality by establishing the two inclusions in turn.
 To show $\Ugroup(r) \subseteq \Tgrouptoo_r \capsp \Bgroup_r$,
 consider an arbitrary class $\alpha \in \Ugroup(r)$.
 We define $\htop = f - r$ and $\hbot = f + r$ and
 note that they are $r$-perturbations of $f$,
 with $\htopinv(a) = \Tspace_r$ and $\hbotinv(a) = \Bspace_r$.
 By definition of the well group,
 $\alpha$ is supported by every $r$-perturbation of $f$, and therefore by
 $\htop$ and by $\hbot$.
 It follows that $\alpha \in \Tgrouptoo_r \capsp \Bgroup_r$.
 To show $\Tgrouptoo_r \capsp \Bgroup_r \subseteq \Ugroup(r)$,
 we consider an arbitrary class $\alpha \in \Tgrouptoo_r \capsp \Bgroup_r$
 and let $h$ be an arbitrary $r$-perturbation of $f$. 
 To finish the proof, we need to show that $\alpha$ is supported by $h$.
 We define $\Cspace_r = h^{-1}[a,\infty) \capsp \Xspace_r$
 and $\Dspace_r = h^{-1}(-\infty,a] \capsp \Xspace_r$. 
 Note that $\Cspace_r \cupsp \Dspace_r = \Xspace_r$
 while $\Cspace_r \capsp \Dspace_r = h^{-1}(a)$.
 Furthermore, the inequality
 $\Maxdist{h}{f} \leq r$ implies that $\Tspace_r \subseteq \Cspace_r$
 and $\Bspace_r \subseteq \Dspace_r$.
 By the Mayer-Vietoris Lemma, $\alpha$ is supported by $h^{-1}(a)$, as required.
\eop

\begin{result}[Remark]
  The Point Formula implies that the well group
  for a Morse function $f$ can change only at critical values
  of the function $f_\Aspace$, where $\Aspace = \{a\}$. 
  In other words, terminal critical values are, in this simple context,
  just ordinary critical values.
  Indeed, if $[r,s]$ is an interval that contains no critical values
  of $f_\Aspace$,
  then there is a deformation retraction $\Xspace_s(f_{\Aspace}) \to \Xspace_r(f_{\Aspace})$
  providing an isomorphism $\Hgroup(\Xspace_s(f_{\Aspace})) \to \Hgroup(\Xspace_r(f_{\Aspace}))$.
  Furthermore, this retraction maps $\Tspace_s$ onto $\Tspace_r$,
  in such a way that that the images of $\Hgroup(\Tspace_r)$
  and $\Hgroup(\Tspace_s)$ in $\Hgroup(\Xspace_s(f_{\Aspace}))$ are identical.
  Similarly, the images of $\Hgroup(\Bspace_r)$ and $\Hgroup(\Bspace_s)$
  in $\Hgroup(\Xspace_s(f_{\Aspace}))$ are identical.
  Hence the well groups $\Ugroup(r)$ and $\Ugroup(s)$ are isomorphic.
\end{result}

\paragraph{The well group of an interlevel set.}
We generalize from a point to an interval, which can be closed,
closed-open, open-closed, or open.
To that end, we define the spaces and maps so that the formula for the well group
is the same in all four cases, and indeed the same as in the Point Formula above.
Assume $a < b$, set $\Aspace = [a,b]$, and let $\Aspace' \subseteq \{a, b\}$.
We thus get $\Xspace_r = \Xspace_r (f_{\Aspace}) = f^{-1}[a-r,b+r]$ and
$\Xspace_r' = \Xspace_r (f_{\Aspace'})$, which is the empty set,
$f^{-1}[b-r,b+r]$, $f^{-1} [a-r, a+r]$, or the union of these two interlevel sets.
Correspondingly, we define the \emph{top} and \emph{bottom interlevel sets}:
\begin{eqnarray*}
      \Tspace_r   &=&          f^{-1} [a+r, b+r] , 
  ~~~ \Tspace_r' ~~\subseteq~~ \{ f^{-1} (a+r), f^{-1} (b+r) \},        \\
      \Bspace_r   &=&          f^{-1} [a-r, b-r] ,
  ~~~ \Bspace_r' ~~\subseteq~~ \{ f^{-1} (a-r), f^{-1} (b-r) \};
\end{eqnarray*}
see Figure \ref{fig:open-interval}.
\begin{figure}[hbt]
 \vspace*{0.1in}
 \centering
 \resizebox{!}{1.8in}{\input{open-interval.pstex_t}}
 \caption{Each vertical strip represents $\Xspace$,
   and the shaded portions mark $(\Cspace_r, \Cspace_r')$ and $(\Tspace_r, \Tspace_r')$
   on the left, $(\Xspace_r, \Xspace_r')$ in the middle,
   and $(\Bspace_r, \Bspace_r')$ and $(\Dspace_r, \Dspace_r')$ on the right.}
 \label{fig:open-interval}
\end{figure}
The pairs $(\Tspace_r, \Tspace_r')$ and $(\Bspace_r, \Bspace_r')$ include
into $(\Xspace_r, \Xspace_r')$ in all four cases.
Still using the notational convention from above, we write $\Tgrouptoo_r$ and $\Bgroup_r$
for the images of  $\Hgroup (\Tspace_r, \Tspace_r')$ and $\Hgroup (\Bspace_r, \Bspace_r')$
in $\Hgroup (\Xspace_r , \Xspace_r')$.
The formula for the well group, $\Ugroup(r) = \Ugroup_{(\Aspace,\Aspace')}(r)$, is then, unsurprisingly:
\begin{result}[Interval Formula]
  $\Ugroup (r) = \Tgrouptoo_r \capsp \Bgroup_r$, for every $r \geq 0$.
\end{result}
\proof
 We give the argument for the most complicated of the four cases,
 when $\Aspace' = \{a,b\}$.
 The proofs of the other three cases are simpler versions of the same argument.
 We may assume $a+r < b-r$, else $\Xspace_r = \Xspace_r'$,
 which implies that all groups in the claimed formula are zero and so we are done.
 To prove the inclusion $\Ugroup(r) \subseteq \Tgrouptoo_r \capsp \Bgroup_r$, 
 we consider the two $r$-perturbations $\htop = f - r$ and $\hbot = f + r$, as before.
 Note that $(\Tspace_r, \Tspace_r') = \htopinv (a,b)$ and 
 $(\Bspace_r, \Bspace_r') = \hbotinv (a,b)$,
 and the desired inclusion follows from the definition of relative well groups.
 To prove $\Tgrouptoo_r \capsp \Bgroup_r \subseteq \Ugroup(r)$, we choose
 an arbitrary class $\alpha \in \Tgrouptoo_r \capsp \Bgroup_r$ and an $r$-perturbation $h$ of $f$.
 Furthermore, we introduce the following pairs of subspaces:
 \begin{eqnarray*}
   \Cspace_r  &=&   h^{-1} [a,\infty) \capsp f^{-1} (-\infty,b+r] ,        \\
   \Cspace_r' &=&  (h^{-1} [a,\infty) \capsp f^{-1} (-\infty,a+r]) \cupsp
                 (h^{-1} [b,\infty) \capsp f^{-1} (-\infty,b+r]),        \\
   \Dspace_r  &=&   h^{-1} (-\infty,b] \capsp f^{-1} [a-r,\infty),         \\
   \Dspace_r' &=&  (h^{-1} (-\infty,a] \capsp f^{-1} [a-r,\infty)) \cupsp
                 (h^{-1} (-\infty,b] \capsp f^{-1} [b-r,\infty));
 \end{eqnarray*}
 see Figure \ref{fig:open-interval} for a depiction of the open case.
 Since $h$ is an $r$-perturbation, we have
 $(\Tspace_r, \Tspace_r') \hookrightarrow (\Cspace_r, \Cspace_r')$
 and similarly $(\Bspace_r, \Bspace_r') \hookrightarrow (\Dspace_r, \Dspace_r')$.
 This implies $\Tgrouptoo_r \subseteq \Cgroup_r$ and $\Bgroup_r \subseteq \Dgroup_r$,
 and therefore $\alpha \in \Cgroup_r \capsp \Dgroup_r$.
 It is easy to see that
 $(\Cspace_r \cupsp \Dspace_r, \Cspace_r' \cupsp \Dspace_r') = (\Xspace_r, \Xspace_r')$,
 and also that
 $(\Cspace_r \capsp \Dspace_r, \Cspace_r' \capsp \Dspace_r') = (h^{-1} (\Aspace), h^{-1} (\Aspace'))$.
 The Mayer-Vietoris Lemma thus implies $\alpha \in (h^{-1} (\Aspace), h^{-1} (\Aspace'))$.
 Since this is true for all $r$-perturbations $h$, we have $\alpha \in \Ugroup(r)$, as required.
\eop

\paragraph{Including intervals.}
We again need some definitions to unify the four cases into one.
Given two intervals $I$ and $J$ of the same type, we say $I$ \emph{includes} into $J$,
denoted as $I \hookrightarrow J$, if $f^{-1} (I)$ includes as a pair in $f^{-1} (J)$.
Unfolding the definition of the four types and assuming $a \leq b \leq c \leq d$, we have
$[b,c] \hookrightarrow [a,d]$,
$[b,d) \hookrightarrow [a,c)$,
$(a,c] \hookrightarrow (b,d]$, and
$(a,d) \hookrightarrow (b,c)$;
compare this with the Mayer-Vietoris diamonds in Figure \ref{fig:pyramid-1}.
Suppose now that we have intervals $I \hookrightarrow J$, both of the same type.
By the Interlevel Set Lemma, there are isomorphisms that take
$\Hgroup_p (f^{-1} (I))$ and $\Hgroup_p (f^{-1} (J))$ to groups
$\Ggroup_p (I)$ and $\Ggroup_p (J)$ in $\Vcal$.
The inclusion induces a map on homology, which composes with these isomorphisms
to give $\gmap : \Ggroup_p (I) \to \Ggroup_p (J)$.
On the other hand, since the two groups are members of $\Vcal$,
there is also a natural map from $\Ggroup_p (I)$ to $\Ggroup_p (J)$,
namely the one that restricts to the identity on the span of their shared vectors
and is zero otherwise.
Not surprisingly, $\gmap$ is exactly that map.
We formalize this claim and give a proof.
\begin{result}[Image Lemma]
  Let $I \hookrightarrow J$ and let $\Ggroup_p (I)$, $\Ggroup_p (J)$ be the corresponding
  $p$-dimensional groups in $\Vcal$.
  Then the image of $\gmap: \Ggroup_p (I) \to \Ggroup_p (J)$
  is a vector space in $\Vcal$,
  and its basis is in bijection with the multiset
  $\Region{p} (I) \capsp \Region{p} (J)$.
\end{result}
\proof
 To restate the lemma, we consider the diagram defined by the homology groups
 of the preimages of the including intervals, $I \hookrightarrow J$,
 and the corresponding vector spaces in $\Vcal$:
 $$
   \begin{array}{ccc}
     \Hgroup_p (f^{-1}(I)) & \stackrel{\hmap}{\longrightarrow} & \Hgroup_p (f^{-1} (J))  \\
     \uparrow              &                                   & \downarrow            \\
     \Ggroup_p (I)         & \stackrel{\gmap}{\longrightarrow} & \Ggroup_p (J)   .                   
   \end{array}
 $$
 The vertical maps are isomorphisms given by the Interlevel Set Lemma.
 The map $\hmap$ is induced by inclusion,
 and $\gmap$ maps a basis vector of $\Ggroup_p (I)$ to the same basis vector
 of $\Ggroup_p (J)$, if it exists, and to zero, otherwise.
 Hence, the basis of $\image{\gmap}$ consists of the vectors that
 are common to the bases of $\Ggroup_p (I)$ and $\Ggroup_p (J)$.
 This lemma states that we can get $\gmap$ by composing $\hmap$ with the two isomorphisms.
 Equivalently, the diagram commutes.
 To prove commutativity, we consider again the zigzag modules drawn as
 monotonic paths in the square; see Figure \ref{fig:pyramid-1}.
 Since $I \hookrightarrow J$, we can find two non-crossing modules,
 one containing $\Hgroup_p (f^{-1} (I))$ and the other containing $\Hgroup_p (f^{-1} (J))$.
 To get a basis for $\image{\hmap}$, we translate intervals from one path to the other,
 keeping only the ones that cover both $\Hgroup_p (f^{-1} (I))$ and $\Hgroup_p (f^{-1} (J))$.
 Further translating these intervals to the hypotenuse gives the
 corresponding points in the persistence diagram.
 These points are precisely the ones shared by
 $\Region{p} (I)$ and $\Region{p} (J)$.
 In other words, $\image{\gmap}$ in $\Vcal$ is isomorphic to $\image{\hmap}$, as desired.
\eop

\begin{figure}[hbt]
  \vspace*{0.1in}
  \centering
  \resizebox{!}{2.8in}{\input{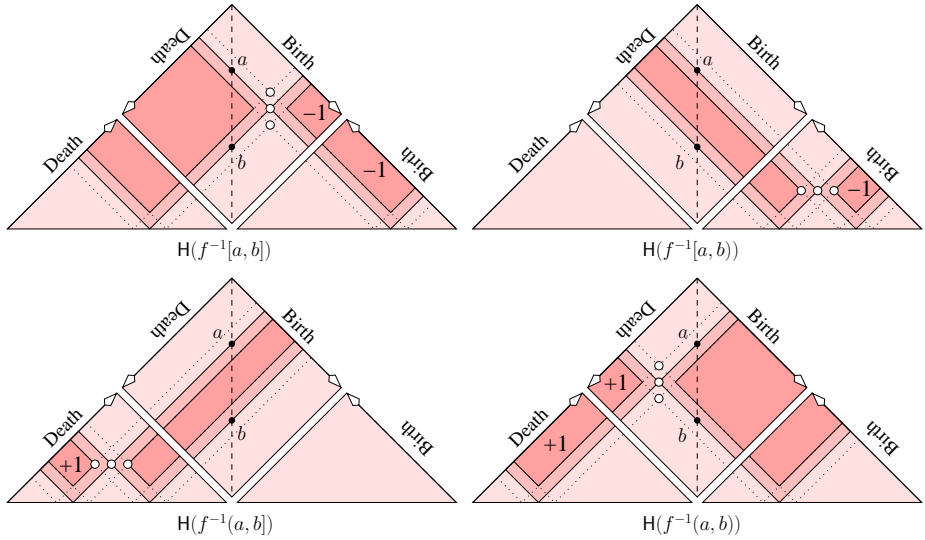}}
  \caption{Reading the robust homology in the four different cases.
    The shaded region gives the basis of $\Hgroup_p (f^{-1} (I))$,
    while the dark shaded region gives the basis of the well subgroup, $\Ugroup_p (r)$.}
  \label{fig:four-robustness}
\end{figure}

\paragraph{Reading robustness.}
The Image Lemma allows us to compute the well groups and the well diagram
associated to a single interval, $I = (\Aspace, \Aspace')$.
The homology of $f^{-1} (I)$ can be read off the persistence diagram of $f$,
as stated in the Interlevel Set Lemma.
Similarly, the homology of $(\Xspace_r, \Xspace_r')$,
where $\Xspace_r = \Xspace_r (f_\Aspace)$ and $\Xspace_r' = \Xspace_r (f_\Aspace')$,
can be read off the same diagram, as we now explain.
By the Interval Formula, the well group for $r$ is the intersection
of the images of the maps
$\tmap_r : \Hgroup_p (\Tspace_r, \Tspace_r') \to \Hgroup_p (\Xspace_r, \Xspace_r')$ and
$\bmap_r : \Hgroup_p (\Bspace_r, \Bspace_r') \to \Hgroup_p (\Xspace_r, \Xspace_r')$
induced by the inclusions.
By the Image Lemma, this intersection corresponds to a pair of rectangles
within the region of $f^{-1} (I)$;
see the intersection between $\Region{p} (I)$ and the dotted rectangles
in Figure \ref{fig:four-robustness}.
In the closed case, this intersection gradually recedes to infinity,
while in the two half-open cases, the intersection disappears when $r$
reaches half the length of the interval.
Correspondingly, the well group shrinks gradually in the closed case,
while it vanishes at or before $r = (b-a)/2$ in the half-open cases.
Similarly, the well group vanishes when $r$ reaches $(b-a)/2$ in the open case.
However, here it vanishes abruptly.
More precisely, the range of the maps $\tmap_r$ and $\bmap_r$,
which is $\Hgroup_p (f^{-1} (a+r, b-r))$,
approaches the homology group of the suspension of the level set at $(a+b)/2$,
when $r$ goes toward $(b-a)/2$, before it suddenly becomes zero when $r$ reaches that limit.

In all four cases, a point contributes to the well group until $r$ reaches a value
at which the shrinking intersection no longer contains the point.
Finding this value of $r$ is easy since both rectangles
shrink uniformly along all of their sides.
Consider for example the case $I = [a,b]$ illustrated by the
upper left design in Figure \ref{fig:four-robustness}.
For a point $(x,y) \in \Ddgm{}{f}$, the value of $r$
at which the point drops out of the relevant region is
\begin{eqnarray*}
  \min \{ b-x , y-b \}  &\mbox{\rm if}&  (x,y) \in \Odgm{}{f} \capsp \lRegion{}[a,b] , \\
  \min \{b-x, y-a \} &\mbox{\rm if}& (x,y) \in \Edgm{}{f} \capsp \lRegion{}[a,b] , \\
  \min \{x-b, a-y \} &\mbox{\rm if}& (x,y) \in \Edgm{}{f} \capsp \rRegion{}[a,b] , \\
  \min \{ x-a , a-y \}  &\mbox{\rm if}&  (x,y) \in \Rdgm{}{f} \capsp \rRegion{}[a,b].
\end{eqnarray*}
The well diagram is the multiset of the values we get from
the points in the persistence diagram.

\paragraph{Measuring the difference.}
We can interpret the rank of the well group as a measure of the similarity
between the image of the map $\tmap_r: (\Tspace_r, \Tspace_r') \to (\Xspace_r, \Xspace_r')$
and the image of the map $\bmap_r: (\Bspace_r, \Bspace_r') \to (\Xspace_r, \Xspace_r')$.
Alternatively, we could use the cokernels of these two maps to
measure their difference.
Indeed, it is not difficult to prove counterparts of the
Image Lemma for cokernels as well as for kernels.
\begin{result}[Co/kernel Lemma]
  Let $I \hookrightarrow J$ and let $\Ggroup_p (I)$, $\Ggroup_p (J)$
  be the corresponding $p$-di\-men\-sional groups in $\Vcal$.
  Then the kernel and cokernel of $\gmap: \Ggroup_p (I) \to \Ggroup_p (J)$
  are vector spaces in $\Vcal$,
  the basis of $\kernel{\gmap}$ is in bijection with $\Region{p}(I) - \Region{p}(J)$,
  and the basis of $\cokernel{\gmap}$ is in bijection with $\Region{p}(J) - \Region{p}(I)$.
\end{result}
To measure the difference, we would therefore take the (algebraic) sum
of the two cokernels.
Consider for example the open case.
By the above lemma, we get a basis of $\cokernel{\tmap_r}$ and $\cokernel{\bmap_r}$
by setting $J = (a+r,b-r)$
and first setting $I$ to $I_1 = (a+r, b+r)$ and second to $I_2 = (a-r, b-r)$.
The basis of the sum, $\cokernel{\tmap_r} + \cokernel{\bmap_r}$,
is in bijection with the union of the two multisets of points,
which is $\Region{p} (J) - \Region{p}(I_1) - \Region{p}(I_2)$.

\section{Discussion}
\label{sec5}

The main contribution of this paper is the introduction of the point calculus
for homology computations of level and interlevel sets.
This comprises interlevel sets defined by closed, half-open, and open intervals,
images, kernels, and cokernels of maps induced by inclusions,
and the robustness of homology as defined by well groups.
The point calculus provides a compact interface to a wealth
of homological information that can be useful to researchers with and without
background in algebraic topology.
For the expert, it provides a compact summary of information that may be used
to formulate conjectures about the topology of spaces and of functions.
For the non-expert, the interface offers an intuitive approach to understand
the topology of datasets that by-passes the introduction of algebraic topology foundations.
It is directly applicable to data in the form of continuous functions,
which is common in medical imaging and in scientific visualization.

We conclude by formulating an open question aimed at casting
light on two- and higher-dimensional notions of robustness.
This paper provides a solution to computing robustness when
$\Yspace = \Rspace$ and perturbations are measured using the $L_{\infty}$-metric,
and \cite{BEKP10} shows that our results also hold for a broader class
of metric function spaces.
In \cite{EMP11}, the authors give an algorithm when $\Xspace$ is an orientable
$2$-manifold, $\Yspace = \Rspace^2$, and $\Aspace$ is a point.
Algorithms for other cases are not yet known.

\section*{Acknowledgements}
The authors thank Vin de Silva for many helpful technical discussions
leading to a strengthening of this paper.

\end{document}